\documentclass[a4paper,twocolumn,11pt,accepted=2025-03-05]{quantumarticle}
\pdfoutput=1
\usepackage[utf8]{inputenc} 
\usepackage[english]{babel} 
\usepackage[T1]{fontenc}
\usepackage{hyperref}
\usepackage{graphicx}  
\usepackage[numbers,sort&compress]{natbib}
\usepackage{subfigure}
\usepackage{amssymb}   
\usepackage{amsmath}
\usepackage{mathtools}
\usepackage{color}
\usepackage[normalem]{ulem}

\usepackage{dcolumn}
\usepackage{bm}
\usepackage{amsthm}

\usepackage{multirow}
\usepackage{algorithm}
\usepackage{algorithmicx}
\usepackage{algpseudocode}

\newtheorem{lemma}{Lemma}
\newtheorem{theorem}{Theorem}

\definecolor{Red}{rgb}{1,0,0}

\def\ket#1{| #1 \rangle}
\def\bra#1{\langle #1 |}

\def\Tr{\operatorname{Tr}}

\def\policy{\operatorname{policy}}
\def\CNOT{\operatorname{CNOT}}
\def\PE{\operatorname{PE}}
\def\tot{\operatorname{tot}}

\begin{document}

\title{Quantum reinforcement learning in continuous action space}

\author{Shaojun Wu}
\affiliation{Institute of  Fundamental and Frontier Sciences, University of Electronic Science and Technology of China, Chengdu, 610051, China}

\author{Shan Jin}
\affiliation{Institute of  Fundamental and Frontier Sciences, University of Electronic Science and Technology of China, Chengdu, 610051, China}

\author{Dingding Wen}
\affiliation{Institute of  Fundamental and Frontier Sciences, University of Electronic Science and Technology of China, Chengdu, 610051, China}

\author{Donghong Han}
\affiliation{Institute of  Fundamental and Frontier Sciences, University of Electronic Science and Technology of China, Chengdu, 610051, China}

\author{Xiaoting Wang$^\dagger$}
\email{xiaoting@uestc.edu.cn}
\affiliation{Institute of  Fundamental and Frontier Sciences, University of Electronic Science and Technology of China, Chengdu, 610051, China}


\begin{abstract}

Quantum reinforcement learning (QRL) is a promising paradigm for near-term quantum devices. While existing QRL methods have shown success in discrete action spaces, extending these techniques to continuous domains is challenging due to the curse of dimensionality introduced by discretization. To overcome this limitation, we introduce a quantum Deep Deterministic Policy Gradient (DDPG) algorithm that efficiently addresses both classical and quantum sequential decision problems in continuous action spaces. Moreover, our approach facilitates single-shot quantum state generation: a one-time optimization produces a model that outputs the control sequence required to drive a fixed initial state to any desired target state. In contrast, conventional quantum control methods demand separate optimization for each target state. We demonstrate the effectiveness of our method through simulations and discuss its potential applications in quantum control.

\end{abstract}

\maketitle

\section{Introduction}

Reinforcement learning (RL)~\cite{sutton2018reinforcement} plays a vital role in machine learning. Unlike supervised and unsupervised learning to find data patterns, the idea of RL is to reduce the original problem into finding a good sequence of decisions leading to an optimized long-term reward, through the interaction between an agent and an environment. This feature makes RL advantageous for solving a wide range of sequential decision problems, including game-playing~\cite{silver2017mastering,mnih2013playing}, e.g., AlphaGo~\cite{silver2016mastering}, robotic control~\cite{peters2003reinforcement,duan2016benchmarking}, quantum error correction~\cite{nautrup2019optimizing,andreasson2019quantum} and quantum control~\cite{palittapongarnpim2017learning,an2019deep,bukov2018reinforcement,niu2019universal,xu2019generalizable,zhang2019does,wauters2020reinforcement}. Typical RL algorithms include Q-learning~\cite{watkins1989,watkins1992q}, Deep Q-Network(DQN)~\cite{mnih2013playing,mnih2015human}, and Deep Deterministic Policy Gradient(DDPG)~\cite{lillicrap2016continuous}. Despite its broad applications, the implementation of RL on classical computers becomes intractable as the problem size grows exponentially, such as the cases from quantum physics. Analogous to quantum computation for conventional computational problems~\cite{nielsen2002quantum}, quantum machine learning has been proposed to solve machine learning problems on quantum computers to potentially gain an exponential or quadratic speedup~\cite{biamonte2017quantum,shor1994algorithms,grover1997quantum,wiebe2012quantum,rebentrost2014quantum,lloyd2018quantum,sarma2019machine,lloyd2014quantum}. 
To date, a variety of quantum reinforcement learning (QRL) algorithms have been developed, leveraging different quantum paradigms to explore potential computational advantages~\cite{meyer2024surveyquantumreinforcementlearning}. Early works on implementing RL on quantum circuits demonstrated a quadratic speedup by leveraging Grover's algorithm~\cite{dong2008quantum,dunjko2016quantum,paparo2014quantum,dunjko2017advances,dunjko2018machine,jerbi2020quantum}. Inspired by deep neural networks, several VQC-based approaches have been proposed, following either value-based~\cite{lockwood2020reinforcement,2020Reinforcement,skolik2021quantum,pmlr-v148-lockwood21a,10313818} or policy-based~\cite{jerbi2021parametrized,pmlr-v202-meyer23a,10313784,Sequeira2023} reinforcement learning frameworks. These methods are designed to be compatible with near-term quantum devices, leveraging parameterized quantum circuits for efficient function approximation. In addition to VQC-based approaches, researchers have also explored QRL methods for fault-tolerant quantum computers, some of which utilize oracle-based access to improve computational efficiency~\cite{dunjko2016quantum,saggio2021experimental,10.1109/SMC.2017.8122616,Cherrat2023,pmlr-v139-wang21w}. Another direction of QRL research is projective simulation, which extends classical reinforcement learning by employing a memory network and quantum random walks, potentially accelerating action selection and offering quantum advantages~\cite{Briegel2012,Melnikov2017,PhysRevX.4.031002,Dunjko_2015,Sriarunothai_2019}. One interesting open question is whether a quantum reinforcement learning(QRL) algorithm can be constructed to guarantee an exponential speedup over its classical counterpart in terms of gate complexity. Besides, another interesting question is how to design the QRL algorithm so that it can efficiently and effectively solve RL problems in \emph{continuous action space}(CAS) without the curse of dimensionality due to discretization, especially the decision problems on quantum systems. In this work, inspired by the classical DDPG algorithm, we propose a quantum DDPG method that can be applied to the quantum state generation in the continuous action space. Specifically, for a fixed state $\ket{s_d}$, we first train the agent to design a parametrized unitary sequence $\{U_a(\bm\theta_t)\}$ that will sequentially drive any state $\ket{s_0}$ into $\ket{s_d}$, where $\ket{s_0}$ can be any state and the action parameters $\bm\theta_t$ take values from a continuous domain and $t=0,\cdots,T$. In our method, the agent's policy and the value function for QRL are constructed from variational quantum neural networks(QNN). The optimal policy function is obtained by continuously optimizing the policy QNN and the value QNN. Once the training is completed, the optimal policy QNN generates the desired sequence $\{U_a(\bm\theta_t)\}$ that transforms any $\ket{s_0}$ to $\ket{s_d}$. Due to the reversibility of unitary operations, the inverse sequence $\{U_a^\dagger(\bm\theta_t)\}$ can be applied to drive $\ket{s_d}$ back to $\ket{s_0}$, where $\ket{s_0}$ can be any state. In particular, if the target state $\ket{s_0}=\ket{s_{\text{unknown}}}$ is unknown but multiple copies are available, one may use the reversed control sequence to efficiently generate $\ket{s_{\text{unknown}}}$ from the fixed initial state $\ket{s_d}$. Hence, our method enables single-shot quantum state generation: a one-time optimization produces a model that outputs the control sequence required to drive a fixed initial state to any desired target state. In contrast, conventional state generation methods typically require separate optimization for each different target state. Moreover, when the target state is unknown, traditional approaches require state tomography to first identify the unknown target state before optimization. In the following, we will first provide a brief introduction to RL and then propose our own QRL algorithm.


\section{Classical reinforcement learning}

\begin{figure}
  \centering
  \includegraphics[width=\columnwidth]{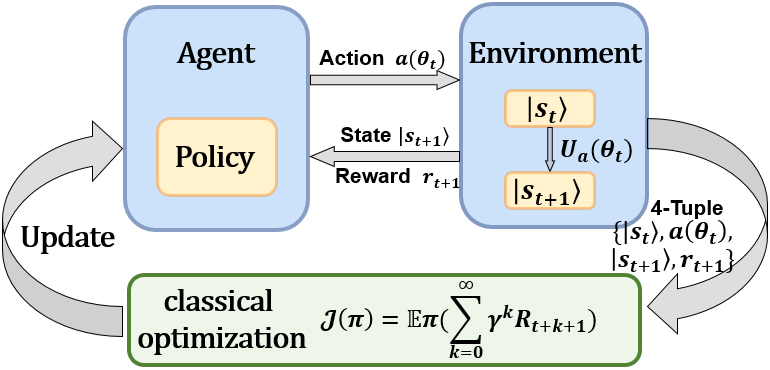}
  \caption{The QRL model. Each iterative step can be described by the following loop: (1) at step $t$, the agent receives $\ket{s_t}$ and generates the  action parameter $\bm\theta_t$ according to the current policy; (2) the agent generates $\ket{s_{t+1}}\equiv U_a(\bm\theta_t)\ket{s_t}$; (3) based on $\ket{s_t}$ and $\ket{s_{t+1}}$, a reward $r_{t+1}$ is calculated and fed back to the agent, together with $\ket{s_{t+1}}$; (4) based on $\ket{s_{t+1}}$ and $r_{t+1}$, the policy is updated and then used to generate $\bm\theta_{t+1}$.}
  \label{fig1}
\end{figure}
In artificial intelligence, an agent is a mathematical abstraction representing an object with learning and decision-making abilities. It interacts with its environment, which includes everything except the agent. The core idea of RL is, through the iterative interactions, the agent learns and selects actions, and the environment responds to these actions, by updating its state and feeding it back to the agent. In the meanwhile, the environment also generates rewards, which are some value-functions the agent aims to maximize over its choice of actions along the sequential interactions~\cite{sutton2018reinforcement}.

Most reinforcement learning problems can be described by a Markov Decision Process (MDP)~\cite{van2012reinforcement,sutton2018reinforcement} with basic elements including a set of states $\mathcal{S}$, a set of actions $\mathcal{A}$ and the reward $\mathcal{R}$. The agent interacts with its environment at each of a sequence of discrete time steps, $t=0,1,\cdots, T$. Each sequence like this generated in RL is called an \emph{episode}. At each time step $t$, the agent receives a representation of the environment's state, denoted by an $N$-dimensional vector $s_t \in \mathcal{S}$, based on which it then chooses an action $a_t\in \mathcal{A}$, resulting the change of the environment's state from $s_t$ to $s_{t+1}$. At the next step, the agent receives the reward $r_{t+1}$ determined by the 3-tuple $(s_{t},a_{t},s_{t+1})$. The aim of the agent is to find a policy $\pi$ that maximizes the cumulative reward $R_t=\sum_{k=0}^T \gamma^k r_{t+k+1}$, where $\gamma$ is a discount factor, $0 \leq \gamma \leq 1$. A large discount factor $\gamma$ means that the agent cares more about future rewards. The policy can be considered as a mapping from $\mathcal{S}$ to $\mathcal{A}$. The update of the policy $\pi$ is achieved by optimizing the value function $Q(s_{t},a_t)\equiv E[R_t|s_{t},a_t]$, i.e., the expectation of $R_t$ under the policy $\pi$. Depending on whether the action space is discrete or continuous, the RL problems can be classified into two categories: DAS(\emph{discrete action space}) and CAS, with different algorithmic design to update the agent's policy. For DAS problems, popular RL algorithms includes Q-learning~\cite{watkins1989}, Sarsa~\cite{rummery1994line}, DQN~\cite{mnih2015human}, etc.; for CAS problems, popular algorithms include Policy Gradient~\cite{NIPS2001_4b86abe4}, DDPG~\cite{lillicrap2016continuous}, etc.

\section{The framework of quantum reinforcement learning}

\begin{figure}
\centering
\includegraphics[width=\columnwidth]{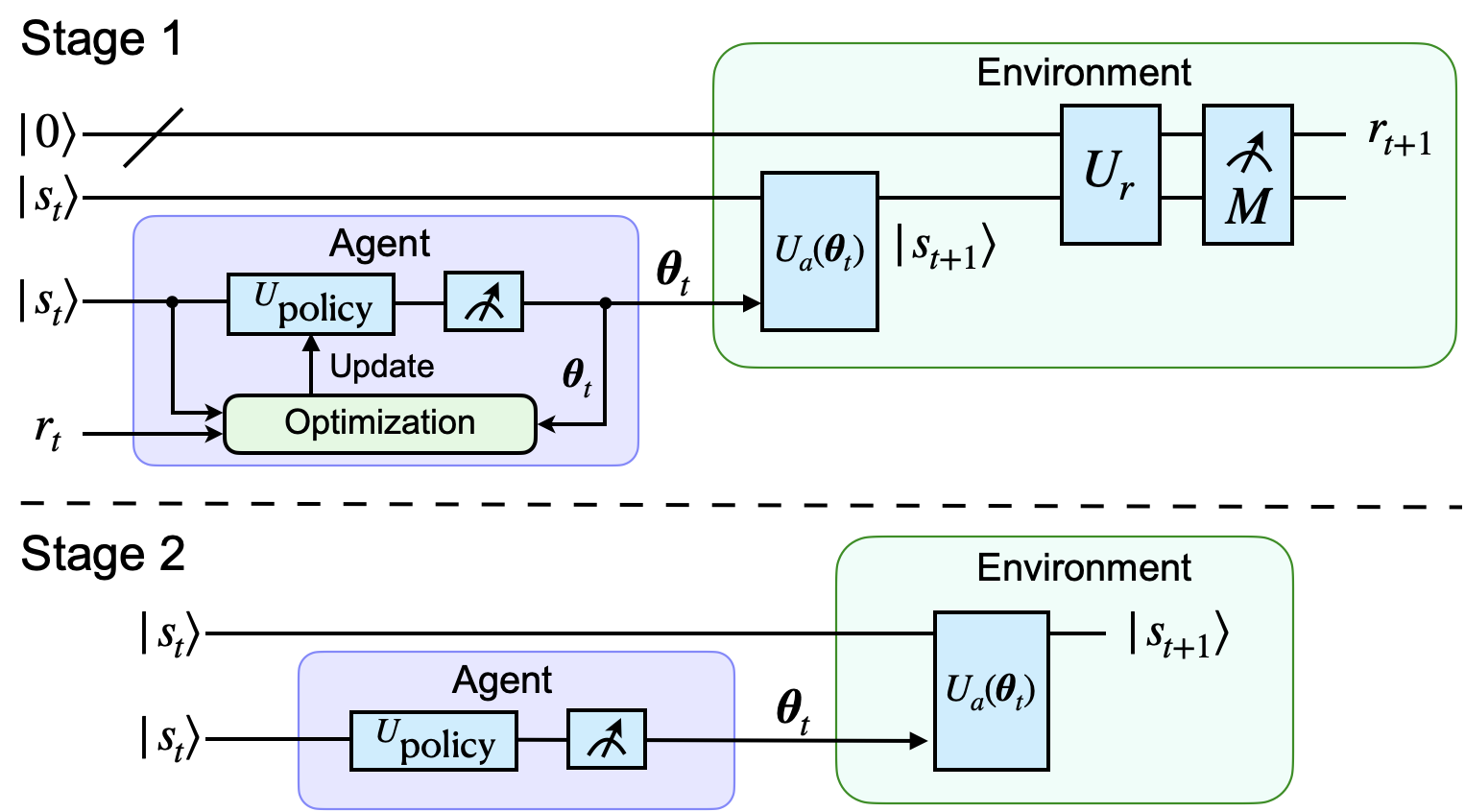}
\caption{The quantum circuit for our QRL framework at each iteration. The entire QRL process includes two stages, so we give the circuit separately. In stage 1, the circuit includes two registers: the reward register, initialized $\ket 0$, and the environment register $\ket {s_t}$. $U_{\policy}$ is generated by the quantum neural network, and determines the action unitary $U_a (\bm\theta_t)$. $U_r$ and $M$ are designed to generate the reward $r_{t+1}$. In stage 2, the circuit has only environment register and does not need to feedback the reward value and update the policy.}
\label{fig2}
\end{figure}

In order to construct a quantum framework that works for both CAS and DAS cases, we present the following QRL model, as shown in Fig.~\ref{fig1}. The essential idea is to map the elements of classical RL into the quantum counterparts. We introduce a quantum `environment' register to represent the environment in RL, and its quantum state $\ket{s_{t}}$ to represent the classical state $s_t$ at time step $t$. Then the action $a(\bm\theta_t)$ can be represented as a parameterized action unitary $U_a (\bm\theta_t)$ on $\ket{s_{t}}$, where $\bm\theta_t$ is the action parameter, which is continuous for CAS case, and takes values from a finite set for DAS case. In order to generate the quantum reward function, by introducing a reward register $\ket{r_t}$, we design the reward unitary $U_r$ and the measurement observable $M$ such that $r_{t+1} \equiv f(\bra{s_t}\bra{0} U_a^\dagger(\bm{\theta}_t) U_r^\dagger M U_r U_a(\bm{\theta}_t) \ket{0}\ket{s_t})$ will match the actual reward defined by the RL problem. Here, $f$ is a function determined by the problem and $\ket{0}$ is the initial state of the reward register. It will be clear in the context of a concrete problem how to design $M$, $U_r$, and $f$ in the correct way, which will be discussed in detail based on the quantum state generate problem and the eigenvalue problem in the following. 

With all RL elements represented as the components of a quantum circuit shown in Fig.~\ref{fig2}, it remains to show how to find the optimal policy $\bm{\theta}_t=\bm{\pi}(\ket{s_t})$ at each time step $t$, such that the iterative sequence $U_{\tot}=U_a(\bm{\theta}_{T})\cdots U_a(\bm{\theta}_1)U_a(\bm{\theta}_0)$ will drive the initial state $\ket{s_0}$ converging to the state $\ket{s_d}$. The entire QRL process can be divided into two stages. In stage 1, we construct the optimal policy $U_{\policy}$ through the agent training, including the policy update and the value-function estimation, which can be realized through the function fitting using QNNs. In stage 2, under the established optimal policy, we can iteratively generate the desired sequence $\{U_a(\bm{\theta}_t)\}$ that will drive the initial state to the target state, and complete the RL task.  


\begin{figure}
\centering
\includegraphics[width=\columnwidth]{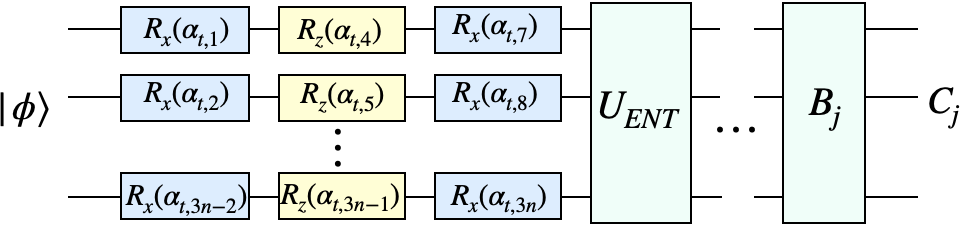}
\caption{Circuit architecture for the VQC. $R_{\beta}(\alpha)\equiv \exp(-i\sigma_{\beta}\alpha/2)$ with $\beta=x,z$. $U_{\textup{ENT}}\equiv \prod_{k=1}^{n-1}\CNOT_{(k,k+1)}$, where $\CNOT_{(k,k+1)}$ denotes the CNOT gate using the $k$-th qubit to control the $(k+1)$-th qubit. $C_j$ is the outcome of the measurement on the observable $B_j$, $j=1,2,\cdots$.}
\label{fig3}
\end{figure}

In our method, in order to solve RL problems in CAS, we utilize QNNs to construct the policy function and the value function. One popular way of implementing a QNN is to use the variational quantum circuit (VQC)~\cite{peruzzo2014variational,farhi2014quantum,lockwood2020reinforcement,benedetti2019parameterized}, whose parameters can be iteratively optimized for the given objective function on classical computers. The VQC circuit of our quantum DDPG algorithm consists of a sequence of unitary $\{D^{(k)}(\bm\alpha)\}$, and is ended by measurements of observables $\{B_j\}$ with $\Tr(B_iB_j)=0$ (Fig.\ref{fig3}), where $B_j=b_{j,1}\otimes b_{j,2} \otimes \cdots \otimes b_{j,n}$ and $b_{j,i}\in \{\sigma_x, \sigma_y, \sigma_z, I\}$. The choice of $\{B_j\}$ are not unique, usually we will choose a set of appropriate $\{B_j\}$ in order to improve the trainability of the quantum neural network. Each $D^{(k)}(\bm\alpha)$ can be chosen to have an identical structure, $D^{(k)}(\bm\alpha)\equiv U_{\textup{ENT}}V^{(k)}(\bm{\alpha})$, where $V^{(k)}(\bm{\alpha})\equiv \otimes_{l=1}^{n}(R_x(\alpha_{k,3l-2})R_z(\alpha_{k,3l-1})R_x(\alpha_{k,3l}))$, $U_{\textup{ENT}}\equiv \prod_{k=1}^{n-1}\CNOT_{(k,k+1)}$ and $\CNOT_{(k,k+1)}$ uses the $k$-th qubit to control the $(k+1)$-th qubit. Here, $R_{x,z}$ are rotations, with $R_{\beta}(\alpha)\equiv \exp(-i\sigma_{\beta}\alpha/2)$, $\beta=x,z$. For the input $\ket {\phi}$, the output of the VQC can be expressed as the expected measurement outcome $C_j\equiv \bra {\phi}D^\dagger (\bm\alpha){B_j} D(\bm\alpha)\ket {\phi}$, based on which the parameter $\bm\alpha$ can then be optimized for the given optimization problem, on a classical computer.

\section{Quantum DDPG algorithm}

For RL problems in CAS, we aim to design QNNs to iteratively construct a sequence of unitary gates that will drive the environment register from the initial state eventually to the target state. This is the essential idea of the quantum DDPG algorithm. Analogous to the classical DDPG algorithm, we make use of the QNNs to construct the desired policy function $\pi_{\bm{\eta}}$ and the value function $Q_{\bm{\omega}}$. Specifically,  the policy-QNN is used to approximate the policy function $\theta_{t,j}\equiv \bra {s_t}D^\dagger (\bm\eta){B_j} D(\bm\eta)\ket {s_t}$ with $\bm\theta_t=(\theta_{t,1},\theta_{t,2},\dots)$, and the Q-QNN is used to approximate the value function $Q(\ket {s_t},\bm\theta_t)\equiv \bra{\bm\theta_t}\bra {s_t}D^\dagger (\bm\omega){B_Q} D(\bm\omega)\ket{s_t}\ket{\bm\theta_t}$. Here, it should be noted that the input of Q-QNN is $\ket{s_t}\ket{\bm\theta_t}$, so we first need encode $\bm\theta_t$ into quantum state $\ket{\bm\theta_t}$ by the amplitude encoding method. For example, we can encode the action parameters as $\ket{\bm\theta_t}=\sum_{j=1}^n \theta_{t,j}\ket{j}$. In order to make the training more stable, two more target networks are included with the same structure as the two main networks~\cite{mnih2013playing,mnih2015human,lillicrap2016continuous}. Therefore, the quantum DDPG uses four QNNs in total: (1) the policy-QNN $\pi_{\bm{\eta}}(\ket{ s_t})$, (2) the Q-QNN $Q_{\bm{\omega}}(\ket {s_t},\bm\theta_t)$, (3) the target-policy $\pi'_{\bm{\eta'}}(\ket {s_t})$ and (4) the target-Q $Q'_{\bm{\omega'}}(\ket {s_t},\bm\theta_t)$. 

The quantum DDPG method contains two stages. In stage 1, the agent training is divided into three parts: (1) experience replay~\cite{lin1992}, (2) updates of the Q-QNN and the policy-QNN, and (3) updates of the target networks. The aim of the experience replay is to prevent the neural network from overfitting. We store the agent's experiences $(\ket {s_t}, \bm\theta_t, r_t,\ket { s_{t+1}})$ in a finite-sized replay buffer $\bm{D}$ at each time step. During the training, we randomly sample a batch of experiences from the replay buffer to update the Q-QNN and the policy-QNN. First, we update the policy-QNN parameters by minimizing the expected return $J=E[Q_{\bm{\omega}}(\ket{s},\bm{\theta})|_{\ket s=\ket{s_i},\bm{\theta}=\pi(\ket {s_i})}]$. Then we update the Q-QNN parameters by minimizing the mean-squared loss $L=\frac{1}{G}\sum_i (y_i-Q_{\bm{\omega}}(\ket{s_i},\bm{\theta}_i))^2$ between the predicted Q-value and the original Q-value, where $y_i=r_i+\bm{\gamma} Q'_{\bm{\omega}'}(\ket{s_{i+1}},\pi'_{\bm{\eta}'}(\ket{s_{i+1}}))$ is the predicted Q-value and calculated by the target-Q networks, $G$ is the size of the batch. Here, we use the gradient descent algorithm AdamOptimizer~\cite{kingma2014adam} to minimize the loss function of these two quantum neural networks. Finally, we update the two target networks using a soft update strategy~\cite{lillicrap2016continuous}: $\bm{\omega}'\gets \tau \bm{\omega}+(1-\tau)\bm{\omega}'$, $\bm{\eta}'\gets \tau \bm{\eta}+(1-\tau)\bm{\eta}'$, where $\tau$ is a parameter with $0<\tau<1$. The entire training process is summarized in Algorithm~\ref{algorithm1}. In stage 2, with the optimal policy-QNN and $T$ iterations, we can construct a sequence of $\{U_a(\bm\theta_t)\}$ and $\ket{s_t}$ for each given initial $\ket{s_0}$, satisfying $\ket{s_T}$ is sufficiently close to the target $\ket {s_d}$.

\begin{algorithm}[H]
\caption{Quantum DDPG algorithm}
\begin{algorithmic}
\State Randomly initialize $Q_{\bm{\omega}}(\ket s,\bm\theta)$ and $\pi_{\bm{\eta}}(\ket s)$;
\State Initialize target $Q'$ and $\pi'$;
\State Initialize replay buffer $\bm{D}$;
\For {episode=1, M}
\State Prepare the initial state $\ket{0,s_0}$;
\For {t=1:$T$}
\State Select the actions: $\bm{\theta}_t=\pi_{\bm{\eta}}(\ket {s_t})$;
\State Apply $U_a(\bm{\theta}_t)$: $\ket{s_{t+1}}=U_a(\bm{\theta}_t)\ket{s_t}$;
\State Apply $U_r$ and $M$ to obtain $r_{t+1}$;
\State Store tuple $(\ket{s_t}, \bm{\theta}_t, r_t, \ket{s_{t+1}})$ in $\bm{D}$;
\State Sample a batch of tuples $(\ket{s_i}, \bm{\theta}_i, r_i, \ket{s_{i+1}})$ from $\bm{D}$;
\State Set $y_i=r_i+\bm{\gamma} Q'_{\bm{\omega}'}(\ket{s_{i+1}},\pi'_{\bm{\eta}'}(\ket{s_{i+1}}))$;
\State Update Q-QNN by minimizing the loss:
\State $L=\frac{1}{G}\sum_i (y_i-Q_{\bm{\omega}}(\ket{s_i},\bm{\theta}_i))^2$;
\State Update the policy-QNN:
\State $\nabla_{\bm{\eta}}J\approx\frac{1}{G}\sum_i \nabla_{\bm{\theta}} Q_{\bm{\omega}}(\ket{s_i},\bm{\theta_i}) \nabla_{\bm{\eta}}\pi_{\bm{\eta}}(\ket {s_i})$;
\State Update the target QNNs:
\State $\bm{\omega}'\gets \tau \bm{\omega}+(1-\tau)\bm{\omega}'; \bm{\eta}'\gets \tau \bm{\eta}+(1-\tau)\bm{\eta}'$.
\EndFor
\EndFor
\end{algorithmic}
\label{algorithm1}
\end{algorithm}

For DAS problems, the above QRL proposal still works if the quantum DDPG design in Fig.~\ref{fig2} is replaced by the quantum DQN design, analogous to the classical DQN algorithm~\cite{mnih2015human}. Compared with the quantum DDPG, the quantum DQN maps states of the environment into the computational basis, rather than into the amplitudes of a quantum register. Moreover, for quantum DQN, only the value function needs to be approximated by the QNN, while the policy function can be described by the greedy algorithm~\cite{sutton2018reinforcement}. Detailed proposals to solve DAS problems using QNNs are presented in~\cite{lockwood2020reinforcement,2020Reinforcement}. It is worthwhile to note that the quantum DQN cannot efficiently solve CAS problems, since the dimensionality problem is inevitable when solving the CAS problems through discretization.

\section{Application of Quantum DDPG in quantum state generation problems}

The quantum state generation problem is a very important ingredient in quantum computation and quantum control. The standard definition of a state generation problem is as follows: given a desired target state that is challenging to prepare, the goal is to design a unitary evolution that drives an easily prepared initial state into the target state. In general, solving this problem requires a case-by-case approach, as any change in the target state necessitates a redesign of the corresponding unitary transformation. An natural question arises: is it possible to develop a unified framework that, for any given target state, automatically generates the appropriate unitary transform to achieve this evolution? The answer is yes. Our proposed method accomplishes precisely this. Our approach begins by designing a process that ensures any given initial state evolves into an easily prepared target state. Then, by swapping the roles of the initial and target states and reversing the process, we construct a general framework that drives the easily prepared initial state into the given target state—regardless of what that target state is.

Specifically, our algorithm has different goals during the training phase (Stage 1) and the execution phase (Stage 2). In stage 1, the purpose of quantum neural network training is to find an optimal policy network that can generate a sequence of $\{U_a(\bm{\theta}_{t})\}$ to drive any given initial state $\ket{s_0}$ to a fixed target state $\ket{s_d}$. Here, the choice of $\ket{s_d}$ is not unique, in fact, in order to realize the measurement of the reward, we will choose some quantum states that are easy to prepare, such as $\ket{0}^{\otimes n}$. In stage 2, leveraging the reversibility of the unitary operation, the inverse sequence $\{U_a^\dagger(\bm{\theta}_{t})\}$ drives the easily prepared initial state $\ket{s_d}$ to the target state $\ket{s_0}$, for any given $\ket{s_0}$. Thus, we have completed the task of state generation. In particular, if the target state is unknown but multiple copies are available, the reversed control sequence can still be applied to efficiently generate it. This highlights the `single-shot' feature of our approach: a one-time optimization yields a model that outputs the desired control sequence to drive a fixed initial state into any desired target state.

The QRL circuit for the state generation is shown in Fig.~\ref{fig2}. First, we define the environment $U_r$, and at each step $t$, we define the reward function $r_{t+1}=f(p_{t+1})$. Given the training target $\ket{s_d}$, we choose an observable $M_{d}\equiv \ket{s_d}\bra{s_d}$. Then we can obtain an estimate of $p_{t+1}\equiv \langle M_{d}\rangle=\bra{s_{t+1}}M_{d}\ket{s_{t+1}}$ through the measurement statistics of $M_d$, with estimation error $\epsilon$. Notice that the number of measurements required to obtain $p_{t+1}$ and $r_{t+1}=f(p_{t+1})$ is independent of the system size $N=2^n$ of the $n$-qubit environment register, and the proof will be shown later in the following. Let $\ket{s _0}=\sum_{k=1}^N \alpha_{0,k}\ket {v_k}$ be the initial state of the environment register, where $\alpha_{0,k}=\langle v_k\ket {s _0}$. At time step $t$, applying $U_{\policy}$ and quantum measurements on $\ket {s_t}$, as shown in Fig.~\ref{fig2}, we obtain the action parameter $\bm\theta_t$, and generate the corresponding action unitary $U_a (\bm{\theta}_t)\equiv U_{\textup{ENT}}V(\bm{\theta}_t)$, where $V(\bm{\theta}_t)$ and $U_{\textup{ENT}}$ are defined as in Fig.~\ref{fig3}. Then we apply $U_a(\bm{\theta}_t)$ and get
\begin{align}\label{eq:action}
\ket{s_{t+1}}=U_a(\bm{\theta}_t)\ket{s_t}= U_{\textup{ENT}}V(\bm{\theta}_t)\ket{s_t}.
\end{align}
Next, by measuring $M_{d}$, we obtain an estimation of 
\begin{align}
p_{t+1}\equiv \bra {s _{t+1}}M_{d}\ket {s _{t+1}}= |\bra{s_{t+1}} s_d\rangle |^2
\end{align}
through $K$ number of measurements. For this state generation problem, we choose the reward $r_{t+1}= 10 p_{t+1}$. If $p_{t+1}\to 1$ as $t\to\infty$, then $\ket{s_{t+1}}$ will converge to $\ket{s_d}$ to complete the RL goal. Notice that the evaluation of $r_{t+1}$ is only needed for the QRL training stage; in the application stage, since $U_{\policy}$ is optimized, measurements are no longer required to estimate $r_{t+1}$. In addition, we will use the same quantum state $\ket{s_t}$ more than once in the algorithm, as shown in Fig.~\ref{fig2}. Since the action parameters $\bm\theta_t$ can be recorded, we can obtain multiple copies of $\ket{s_t}$ by repeating preparation.

\begin{figure}
\centering
\includegraphics[width=\columnwidth]{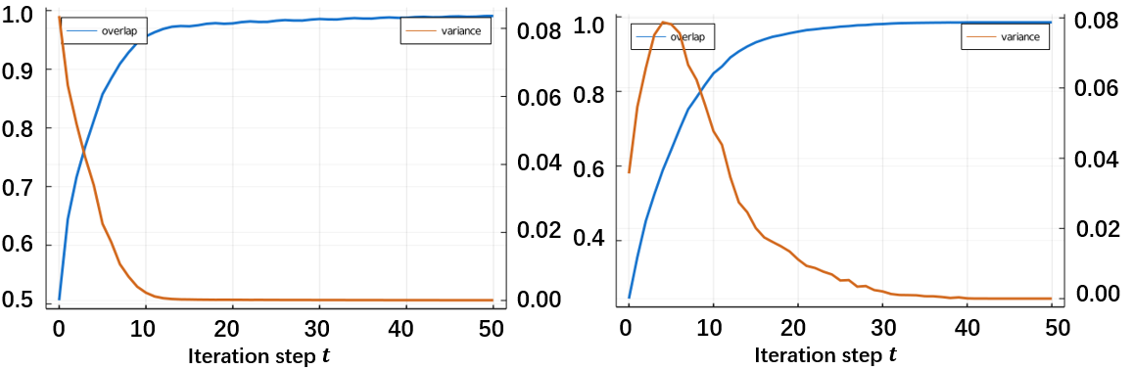}
\caption{Simulation results for quantum state generation problem of the one-qubit and the two-qubit Hamiltonian by quantum DDPG. For $1000$ different initial $\ket{s_0}$, we plot how the average $\bar p_t$ and the variance $\Delta(p_{t})$ change with the iteration step $t$. For the one-qubit case, at $t=50$, $\bar p_{50} \ge 0.99$ and $\Delta(p_{50})\leq 4.47\times 10^{-5}$. For the two-qubit case, at $t=50$, $\bar p_{50} \ge 0.98$ and $\Delta(p_{50})\leq 4.04\times 10^{-7}$.}
\label{fig6}
\end{figure}

To verify the effectiveness of our method, we study the state generation problem for one-qubit and two-qubit cases. In stage 1, we apply the quantum DDPG algorithm to update the policy until we obtain the optimal $U_{\policy}$. Since the interaction between agent and environment is influenced by noise, we randomly choose the error $\epsilon\sim\mathcal{N}(p_t,\epsilon^2)$ on the reward value in the simulation. In stage 2, based on $U_{\policy}$ and $\ket{s_0}$, we generate a sequence of $\{U_a(\bm\theta_t)\}$ and the corresponding $\ket{s_{50}}= U_a(\bm{\theta}_{49})\cdots U_a(\bm{\theta}_0)\ket{s_0}$, and record the overlap $p_t$ at each $t$.

Specifically, for the one-qubit case, the training target state is chosen to be $\ket 1$. In stage 1, the size of the replay buffer is set as $1000$, the size of the batch is set as $16$, and the other parameters are set as $\gamma=0.9$, $\tau=0.001$. In addition, the quantum registers for the policy-QNN and the value-QNN are both comprised of three qubits, and the depth of the QNN circuits is one. For the policy-QNN, in order to increase the nonlinearity of the QNN, we use auxiliary qubits. Specifically, we add two ancilla qubits initialized in $\ket{00}$ together with $\ket{s_t}$ as the input of the policy network. Usually, the size of the auxiliary qubits is one to two times the number of qubits in the environment state $\ket{s_t}$. For the value-QNN, we encode $\bm\theta_t$ into $\ket{\bm\theta_t}$ and use it together with $\ket{s_t}$ as the inputs of the value-QNN. Then we perform the training process. After the training stage, we randomly select 1000 different initial states $\ket{s_0}$ to test the performance of the policy $U_{\policy}$. In Fig.~\ref{fig6}, we can see that almost all final states $\ket{s_{50}}$ are sufficiently close to the target $\ket {s_d}$, with $p_{50} \ge 0.99$ and $\Delta(p_{50})\leq 4.67\times 10^{-5}$ at $t=50$. For the two-qubit case, the target state $\ket {s_d}$ is chosen to be the maximum entangled state $(\ket {00} +\ket {11})/\sqrt{2}$. As the number of qubits of the environment state increases, the training of the neural network will become more difficult. In our simulation, using the original quantum neural network training can not get satisfactory results. In order to address this problem, we apply the universal quantum neural network proposed in Ref.~\cite{hou2021universal}, which introduced a classical activation function and a weight matrix on the basis of VQC. The QNNs with this structure has the ability to approximate arbitrary functions. Also, we set the size of the replay buffer as $10000$, the size of the batch as $128$, and other parameters as $\gamma=0.9$, $\tau=0.005$. Here, the quantum registers to implement these two QNNs both contain six qubits, and the depth of both QNN circuits is three. Simulation results in Fig.~\ref{fig6} show that $\bar p_{50} \ge 0.98$ and $\Delta(p_{50})\leq 4.04\times 10^{-7}$ at $t=50$. Notice that for both one-qubit and two-qubit cases, a one-shot optimization is sufficient to find the optimal policy $U_{\policy}$ through QNN learning. Once the QRL model is constructed, for each $\ket {s_0}$, it can efficiently generate the desired control sequence $\{U_a (\bm{\theta}_{t})\}_{t=0}^{49}$ to drive $\ket {s_0}$ to $\ket {s_d}$. If we take $\ket {s_d}$ as the initial state, then the reversed sequence $\{U_a^\dag (\bm{\theta}_{t})\}_{t=49}^{0}$ will drive $\ket {s_d}$ to different $\ket {s_0}$. Thus, we have completed the task of any given state generation. It is worthwhile to note that it is not necessary for $\ket {s_0}$ to be known to make our model work: even if $\ket {s_0}$ is unknown, as long as we have sufficient number of identical copies of $\ket {s_0}$, our model is still able to output the desired sequence that will drive $\ket {s_d}$ to $\ket {s_0}$.

\section{Quantum DDPG in solving the eigenvalue problem}

\begin{figure}
\centering
\includegraphics[width=\columnwidth]{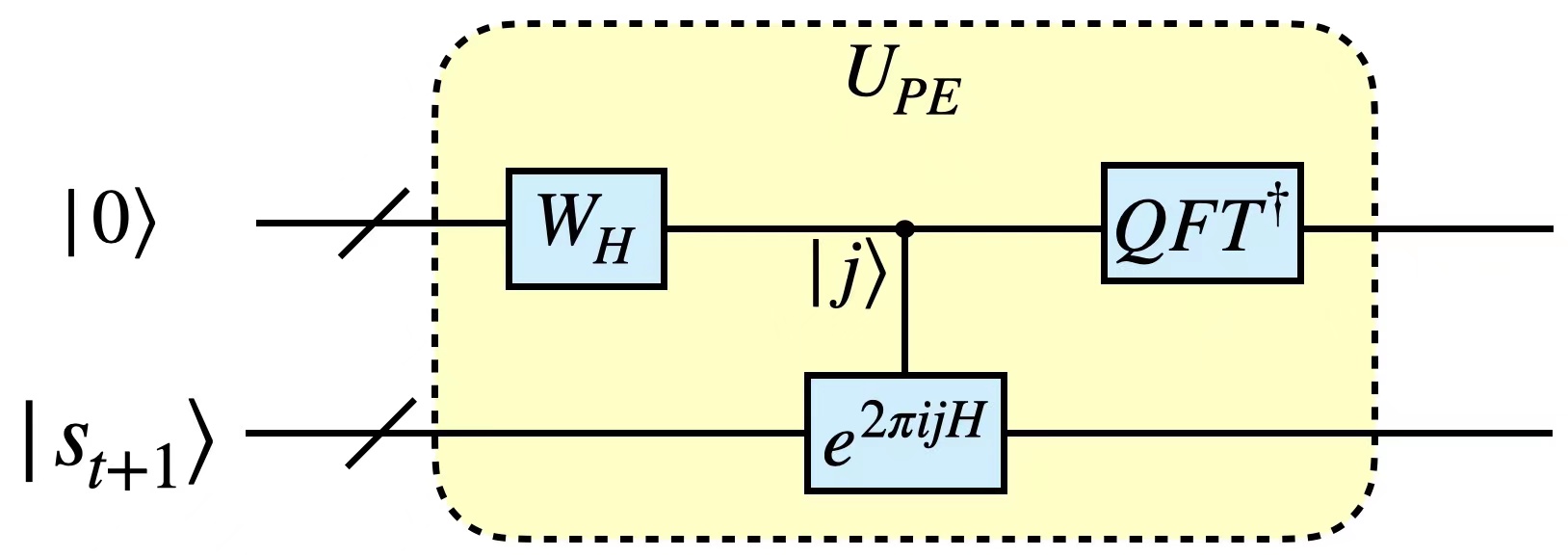}
\caption{The quantum phase estimation circuit $U_{\PE}$.}
\label{fig4}
\end{figure}

In quantum complexity theory, the Quantum Merlin Arthur(QMA) is the set of all decision problems that can be verified by quantum computers in polynomial time~\cite{aharonov2002quantum,watrous2008quantum,gharibian2014quantum}. One typical QMA-complete problem is the $k$-local Hamiltonian problem, which is to find the ground energy of a $k$-local Hamiltonian $H$ with $k \ge 2$~\cite{kempe2006complexity}. Essentially, this problem is an eigenvalue problem for a given physical Hamiltonian, and one way of solving it on a quantum computer is to use the variational quantum eigensolver (VQE) algorithm~\cite{peruzzo2014variational}. Here, we present an alternative method, formulating the $k$-local Hamiltonian problem as a reinforcement learning problem in CAS and applying our quantum DDPG algorithm to it. Specifically, let $H$ be the Hamiltonian defined on $N$ dimensional quantum system $E$, and $H$ is a sparse matrix that can be efficiently constructed. Assuming an unknown eigenvalue $\lambda_0$ of $H$ is located in a neighborhood of $\bar\lambda$, i.e. $\lambda_0\in \delta(\bar\lambda)\equiv[\bar\lambda-\delta,\bar\lambda+\delta]$, we aim to find out $\lambda_0$ and its corresponding eigenvector $\ket{u_0}$. To implement the QRL circuit in Fig.~\ref{fig2} for the eigenvalue problem, we choose $U_r$ as the quantum phase estimation $U_{\PE}$ shown in Fig.~\ref{fig4}. The role of $U_{\PE}$ together with the subsequent measurement is to map the input state $\ket{s_{t+1}}$ into the desired eigenstate with certain probability. Specifically, the reward function $r_{t+1}$ can be defined as the overlap between the $(t+1)$-th states with $\ket{u_0}$: $r_{t+1}\equiv 10|\bra{s_{t+1}} u_0\rangle |^2$. Let $\ket 0$ and $\ket{s _0}=\sum_{k=1}^N \alpha_{0,k}\ket {u_k}$ be the initial states of the reward register and the $n$-qubit environment register, where $n=\log N$ and $\alpha_{0,k}=\langle u_k\ket {s _0}$. At the time step $t$, applying $U_{\policy}$ and quantum measurements to $\ket {s_t}$, we obtain the action parameter $\bm\theta_t$, and the next state is $\ket{s_{t+1}}=U_a(\bm{\theta}_t)\ket{s_t}$.

Then, by applying $U_{\PE}$, we obtain
\begin{align}\label{eq:upe}
 U_{\PE}\ket 0\ket {s _{t+1}}=\sum_{k=1}^N \alpha_{t+1,k}\ket{{\lambda}_k}\ket{u_{k}},
\end{align}
where $\ket {u_k}$ is the eigenvector corresponding to the eigenvalue $\lambda_k$. Here, the input state of the eigenstate register is in a superposition of different eigenstates, so the output state becomes an entangled state between the eigenvalue phase register and the the eigenstate register. Therefore, in the next step, we only need to measure the eigenvalue phase register on the computational basis to obtain $|\bra{s_{t+1}} u_0\rangle |^2$. Specifically, by measuring the eigenvalue phase register, we obtain the outcome $\lambda_0$ with probability
\begin{align}\label{eq:overlap}
p_{t+1}\equiv  |\bra{s_{t+1}} u_0\rangle |^2 = |\alpha_{t+1,0} |^2
\end{align}
which can be estimated by the frequency of obtaining $\lambda_0$ in $K$ number of measurements. The reward can be written as $r_{t+1}=10p_{t+1}$.

\begin{figure}
\centering
\includegraphics[width=\columnwidth]{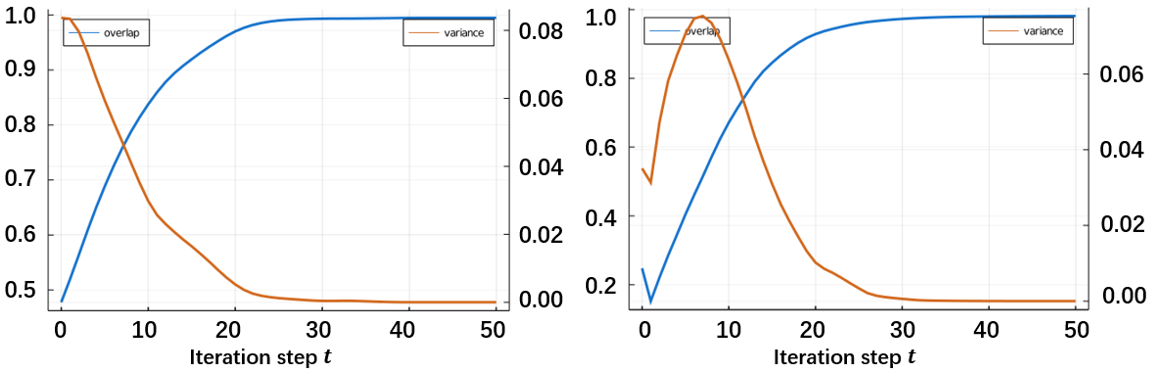}
\caption{Simulation results for solving the eigenvalue problem of the one-qubit and the two-qubit Hamiltonian by quantum DDPG. For $1000$ different initial $\ket{s_0}$, we plot how the average $\bar p_t$ and the variance $\Delta(p_{t})$ change with the iteration step $t$. For the one-qubit case, at $t=50$, $\bar p_{50} \ge 0.99$ and $\Delta(p_{50})\leq 1.98\times 10^{-7}$. For the two-qubit case, at $t=50$, $\bar p_{50} \ge 0.98$ and $\Delta(p_{50})\leq 3.01\times 10^{-6}$.}
\label{fig5}
\end{figure}


Next, as numerical demonstration, we apply this method to the Heisenberg model with $n=1,2$. In stage 1, the parameter settings of the quantum neural networks are the same as in the cases of $n=1,2$ in the state generation problem. After the training in stage 1, we randomly select 1000 different initial states $\ket{s_0}$ to test the policy $U_{\policy}$. For the one-qubit case, we choose the Hamiltonian as $H\equiv \frac{1}{4}(0.13 \sigma_x+0.28 \sigma_y+0.95 \sigma_z+I)$, where the three coupling constants are randomly generated.  One can see from Fig.~\ref{fig5} that as $t$ increases, trajectories $\{\ket{s_t}\}$ with different initial $\ket{s_0}$ all converge to the ground state $\ket{u_0}$, with $\bar p_{50} \ge 0.99$ and $\Delta({p_{50}})\leq 1.98\times 10^{-7}$ at $t=50$. Next, we consider a two-qubit Hamiltonian model $H=\frac{1}{2}(\sigma_x\otimes\sigma_x
+\sigma_y\otimes\sigma_y+\sigma_z\otimes\sigma_z)+0.25\sigma_x\otimes I$, and use the quantum DDPG algorithm to find an energy state of it. The simulation result is shown in Fig.~\ref{fig5} that as $t$ increases, trajectories with different initial $\ket{s_0}$ all converge to the ground state of $H$, with $\bar p_{50} \ge 0.98$ and $\Delta({p_{50}})\leq 3.01\times 10^{-6}$ at $t=50$.  However, we are not satisfied with the accuracy rate of 0.98, so we can use the final state $\ket{s_{50}}$ as the input of the method of Ref.~\cite{PhysRevLett.83.5162} to obtain an exact state.

Again, analogous to the state generation problem, our QRL method demonstrates an advantage over the conventional optimal control~\cite{khaneja2005optimal}: the optimal $U_{\policy}$ generated through a one-shot optimization is useful for any given initial state, while a different optimization is required for each different initial state in optimal control or VQE. 

\section{Complexity analysis}

Analogous to the VQE algorithm, our quantum DDPG algorithm is essentially a quantum-classical hybrid optimization. Similar to other discussion, we will mainly focus on the quantum circuit complexity of our algorithm, including the circuit gate complexity and the measurement complexity. In our method, quantum measurements are required to obtain both the reward value and the action parameters.

Let $B$ be an observable with $B \ket{u_j}=\lambda_j  \ket{u_j}$, for an $n$-qubit system, $N=2^n$. Assuming the system is in the state $\ket{\psi}=\sum_{j=1}^N \alpha_j \ket{u_j}$, the measurement of $B$ in $\ket{\psi}$ will generate the measurement outcome $m$, with probability distribution $P(m=\lambda_j)=|\langle u_j \ket{\psi}|^2=|\alpha_j|^2$. Notice that $m$ and $B$ have the same expectation value and the variance: $\mathbb{E}(m) =\langle B\rangle$ and $\text{Var}(m)=\langle B^2\rangle-\langle B\rangle^2$. Then we have the following well-known result in probability theory:

\begin{lemma}[Chebyshev's inequality]
\label{lemma1}
Let $X$ is a random variable with expected value $\mu$ and variance $\sigma^2$. For any real number $k>0$, $P(|X-\mu|\ge k\sigma) \leq \frac{1}{k^2}$.
\end{lemma}

Based on Lemma $1$, we further derive the following relationship between the measurement precision error $\epsilon$ and the number of measurements $K$ to reach that precision.

\begin{theorem}
\label{thm-measure}
Let $B$ be an observable with $B=b_1\otimes b_2 \otimes \cdots \otimes b_n$, where $b_i\in \{\sigma_x, \sigma_y, \sigma_z, I\}$. We implemented measurements for $K$ times to obtain the sample average $\overline {m}_K$ to estimate the expected value $\langle B\rangle$. Then, the  probability of the difference between $\overline {m}_K$ and $\langle B\rangle$ lager than $\epsilon$ is given by
\begin{align}\label{eq:L2}
P(|\overline {m}_K-\langle B\rangle|\ge \epsilon) \leq \frac{1}{K \epsilon^2}.
\end{align}
\end{theorem}

\begin{proof}

After $K$ number of measurements on $B$, we obtain a sample of measurement outcomes, $m_1, m_2, \cdots, m_K$, with sample mean $\overline {m}_K\equiv \frac{1}{K} \sum_{i=1}^K m_i$ and $\{m_i\}$ to be independent and identically distributed. Let the expectation and the variance of $m_i$ be $\mu$ and $\sigma^2$. Due to the weak law of large numbers, we have $\overline {m}_K \to \mu$ for $n\to\infty$. Then the expectation of the sample mean is $\mathbb{E}(\overline {m}_K)=\mu=\langle B\rangle$ and the variance is $\text{Var}(\overline {m}_K)=\frac{\sigma^2}{K}$.
Choosing $k= \frac{\epsilon\sqrt{K}}{\sigma}$ in Lemma~\ref{lemma1} and using Chebyshev's inequality result in $P(|\overline {m}_K-\langle B\rangle|\ge \epsilon) \leq \frac{\sigma^2}{K \epsilon^2}$. 

For the observable $B=b_1\otimes b_2 \otimes \cdots \otimes b_n$ and $b_i\in \{\sigma_x, \sigma_y, \sigma_z, I\}$, we have $\lambda_i=\pm1$ and $\langle B^2\rangle$ is $1$. Further, we have $0\leq \langle B\rangle^2 \leq 1$ and $0\leq \langle B^2\rangle-\langle B\rangle^2 \leq 1$. Hence, $0\leq \sigma \leq 1$, and we find $P(|\overline {m}_K-\langle B\rangle|\ge \epsilon) \leq \frac{1}{K \epsilon^2}$. 
\end{proof}
Now we apply Theorem~\ref{thm-measure} to analyze the measurement complexity of our algorithm. For a given estimation error $\epsilon$, according to Theorem~\ref{thm-measure}, choosing $K=\mathcal{O}(\frac{1}{\epsilon^2})$ will make $\overline {m}_K$ converge to $\langle B\rangle$ with high probability. Next, we analyze the gate complexity of our algorithm. During a single iteration at $t$ of stage 1, denoting the gate complexities of $U_a(\bm{\theta_t})$, $U_{\policy}$(policy-QNN) and $U_{Q}$(Q-QNN) as $C_a$, $C_p$ and $C_q$, we can see that the number of parameters $\bm{\theta_t}$ in $U_a(\bm{\theta_t})$ is at most equal to $C_a$. Hence, in a single iteration at $t$ of stage 1, the gate complexity is $\mathcal{O}(\frac{C_a C_p+C_a^2 t+C_q}{\epsilon^2})$. Analogously, in a single iteration at $t$ of stage 2, the gate complexity is $\mathcal{O}(C_a t)+\mathcal{O}(\frac{C_p+C_a t}{\epsilon^2} C_a)$; hence, the total gate complexity of stage 2 is $\mathcal{O}(\frac{T^2 C_a^2+C_a C_p}{\epsilon^2})$.

\section{Concluding discussion}
In this work, inspired by the classical DDPG algorithm, we have proposed a quantum DDPG algorithm that can solve both CAS and DAS reinforcement learning problems. For CAS tasks, our quantum DDPG algorithm encodes the environment state into the quantum state amplitude to avoid the dimensionality disaster due to discretization. As a useful application, our method can be used to solve the state generation problem. A distinguishing feature of our method is that, for each target state, it only requires a one-shot optimization to construct the QRL model that is able to efficiently output the desired control sequence driving the initial state to the target state. In comparison, in conventional quantum control methods, different optimizations are required for each different target state. Simulation results for one-qubit and two-qubit quantum systems demonstrate that, our QRL method can be used for any given state generation and eigenstate preparation for a given Hamiltonian. We have also analyzed the complexity of our proposal in terms of the QNN circuit complexity and the measurement complexity.

\section*{Acknowledgments}
This research was supported by the National Natural Science Foundation of China (Grant No.92265208) and the National Key R\&D Program of China (Grant No.2018YFA0306703). We also thank Xiaokai Hou, Yuhan Huang, and Qingyu Li for helpful and inspiring discussions.


\bibliographystyle{unsrtnat}
\bibliography{ref}

\onecolumn\newpage
\appendix

\section{Classical Reinforcement Learning}

\begin{figure}[h]
\centering
\includegraphics[width = 3.5in]{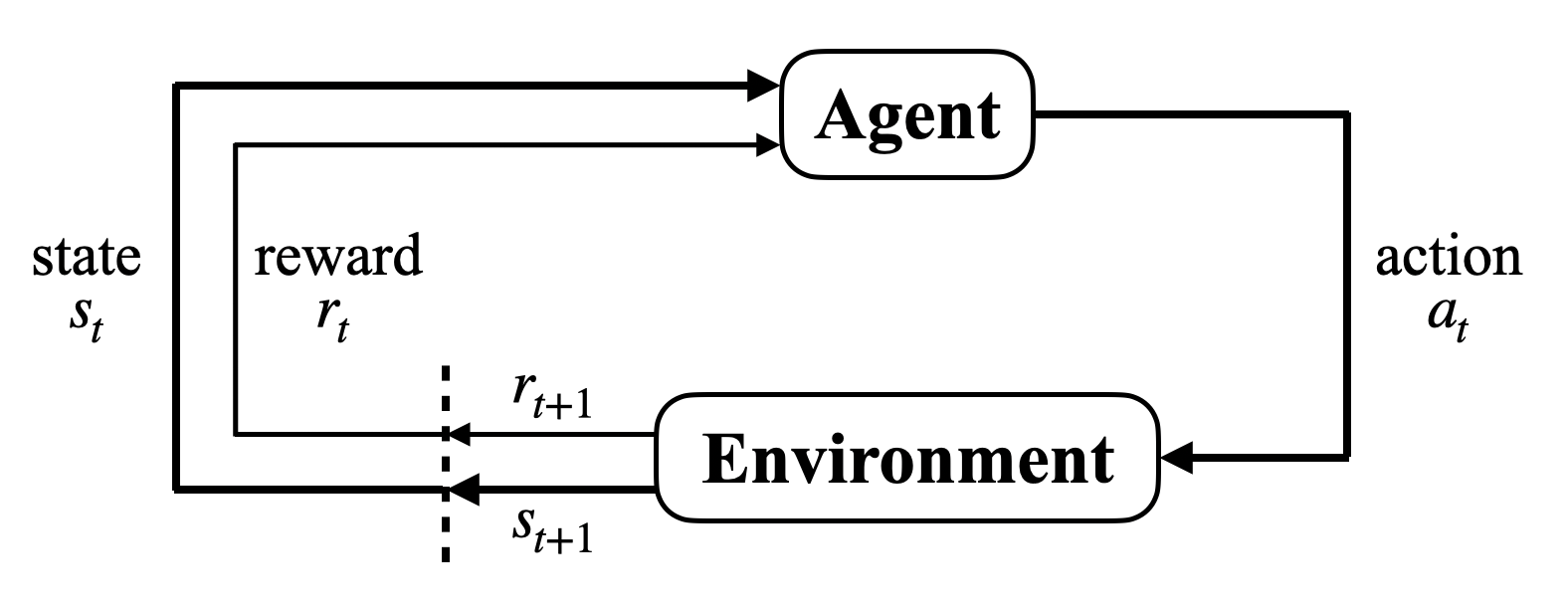}
\caption{An illustration of the standard RL method. }
\label{fig7}
\end{figure}

Here, we will give a detailed introduction to the basic concepts of reinforcement learning for beginners to learn. In RL, the basic elements include a set of states $\mathcal{S}$, a set of actions $\mathcal{A}$, the reward $\mathcal{R}$. The model of reinforcement learning is shown in Fig.~\ref{fig7}~\cite{sutton2018reinforcement}, the agent and the environment interact continually. In a step, the agent receives an observation $s_t$, then chooses an action $a_t$. Next, the agent performs an action $a_t$, and the environment move to next state $s_{t+1}$ and emits a reward $r_{t+1}$. At the next step, the agent receives the reward $r_{t+1}$ determined by the 3-tuple $(s_{t},a_{t},s_{t+1})$. Given an initial state $s_0$, the agent-environment interactions will generate the following sequence: $s_0, a_0,r_1,s_1,a_1,r_2,\cdots$. Such a sequence is called an \emph{episode} in RL. Next, we define the following three key elements of RL:

(1) Policy 

The policy can be considered as a mapping from $\mathcal{S}$ to $\mathcal{A}$, which sets the rules on how to choose the action based on the environment's state. Such policy is determined by certain optimization objective, such as maximizing the cumulative reward. A policy can be either deterministic or stochastic. A deterministic policy is characterized by a function $a=\pi(s)$, meaning that under the same policy, at time step $t$, the action $a_t$ is uniquely determined by the current environment's state $s_t$. Given the state $s$, we define the stochastic policy, $\pi_\theta(a|s)\equiv P[a|s,\theta]$ as the probability of choosing the random action $a$, where $\theta$ is parameter charactering the distribution of $P[a|s,\theta]$.

(2) Cumulative reward 

As mentioned above, at time step $t$, the policy goal of the agent is to maximize the cumulative reward it receives in the long run. At each step $t$, if we define the accumulative reward as $R_t=\sum_{k=0}^{\infty} r_{t+k+1}$, it may not be convergent and becomes ill-defined; alternatively, we can introduce a discount factor $\gamma( 0 \leq \bm{\gamma} \leq 1)$ and define the cumulative reward as $R_t=\sum_{k=0}^\infty \gamma^k r_{t+k+1}$. The larger the discount factor, the longer time span of future rewards we will consider to determine the current-state policy. At time step $t$, the reward $r_t$ determines the immediate return, and the cumulative reward $R_t$ determines the long-term return.

(3) Value function 

Notice that when $a_t$ or $s_t$ is stochastic, $r_t$ and $R_t$ are also stochastic. Hence, we further define the value function $Q$ to be the expectation of the cumulative reward, $Q(s,a)\equiv E[R_t|s,a]$, under the policy $\pi$. The goal of RL is to find the optimal policy that maximizes the value function $Q$.

RL problems can be classified into two categories: discrete-action-space(DAS) problems and continuous-action-space (CAS) problems. In a DAS problem, the agent chooses the action from a finite set $\{a_k\}$, $k=1,\cdots, l$. For example, in the Pong game~\cite{mnih2015human}, the action set for moving the paddle is \{up, down\}. In a CAS problem, the action can be parametrized as a real-valued vector\cite{masson2015reinforcement}. In the CartPole environment~\cite{doya2000reinforcement}, the action is the thrust and can be parametrized as a continuous variable $\theta \in [-1,1]$. For DAS problems, popular RL algorithms include Q-learning~\cite{watkins1989}, Sarsa~\cite{rummery1994line}, Deep Q-learning Network(DQN)~\cite{mnih2015human}, etc.; for CAS problems, popular algorithms include Policy Gradient~\cite{NIPS2001_4b86abe4}, Deep Deterministic Policy Gradient(DDPG)~\cite{lillicrap2016continuous}, etc. 

Notice that the DQN algorithm is only efficient when solving problems with small DAS. It quickly becomes inefficient and intractable when the size of the DAS becomes large. Hence, although a CAS problem can be converted into a DAS problem through discretization, the DQN algorithm will not work in solving the converted DAS problem, if we require high discretization accuracy. For CAS problems, it is better to use a CAS algorithm, such as DDPG.

\section{Discrete Action Space Algorithms}

Q-learning is a milestone in reinforcement learning algorithms. The main idea of the algorithm is to construct a Q-table of state-action pairs to store the Q-values, and then the agent chooses the action that can obtain the largest cumulative reward based on the Q-values.

In the Q-learning algorithm, an immediate reward matrix $R$ can be constructed to represent the reward value from state $s_t$ to the next state $s_{t+1}$. The Q-value is calculated based on the matrix $R$, and it is updated by the following formula~\cite{sutton2018reinforcement}:
\begin{align}\label{eq1}
\begin{split}
Q(s_t,a_t)&\longleftarrow Q(s_t,a_t)\\
&+\alpha[r_t+\bm{\gamma} \max_{a_{t+1}}Q(s_{t+1},a_{t+1})-Q(s_t,a_t)]
\end{split}
\end{align}
where $\bm{\gamma}$ is the discount factor, $\alpha$ is the learning rate that determines how much the newly learned value of $Q$ will override the old value. By training the agent, the Q-value will gradually converge to the optimal Q-value.

The Q-learning is only suitable for storing action-state pairs which are low-dimensional and discrete. In large-space tasks,  the corresponding Q-table could become extremely large, causing the RL problem intractable to solve, which is known as the curse of dimensionality. The DQN algorithm takes the advantage of deep learning technique to solve the RL problems. Specifically, it introduces a neural network defined as Q-network $Q(s_t,a_t;\bm{\omega})$, whose function is similar to the Q-table in approximating the value function. The input of the Q-network is the current state $s_t$, and the output is the Q-value. The $\epsilon$-greedy strategy is used to choose the value of $a$ according to the following probability distribution~\cite{sutton2018reinforcement}: 
\begin{align}\label{eq2}
a = 
\begin{cases} 
\arg \max_{a}Q(a),  & \text{ with probability}\  1-\epsilon; \\
\text{a random action}, \, &\text{ with probability}\  \epsilon.
\end{cases}
\end{align}

In order to stabilize the training, the DQN algorithm uses two tricks: experience replay and a target network. The method of experience replay is to use a replay buffer to store the experienced data and to sample some data from the replay buffer at each step $t$ to update the parameters in the neural network. The DQN algorithm introduces a target-Q network $Q(s_{t+1},a_{t+1};{\bm{\omega}}')$ which is a copy of the Q-network. The input of $Q(s_{t+1},a_{t+1};{\bm{\omega}}')$ is $s_{t+1}$ and the output is $Q(s_{t+1},a_{t+1})$. However, the target-Q network parameters are updated using Q-network parameters at every $m$ steps, where $m$ is a constant. The DQN algorithm updates the Q-network by reducing the value of the loss function $L(\bm{\omega})=E[((r_t+\bm{\gamma} \max_{a_{t+1}}Q'(s_{t+1},a_{t+1};{\bm{\omega}}'))-Q(s_t,a_t;\bm{\omega}))^2]$.

\section{Continuous Action Space Algorithms}
\begin{figure}
\centering
\includegraphics[width = 2.4in]{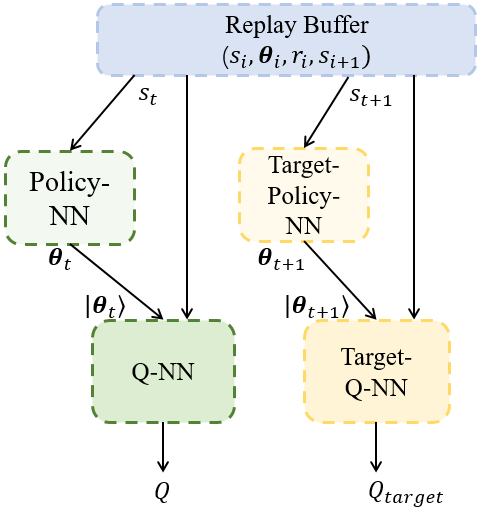}
\caption{The neural network structure of DDPG Algorithm.}
\label{fig9}
\end{figure}
For tasks in continuous action space, we usually use the DDPG algorithm. The DDPG algorithm make use of the neural network to construct the desired policy function $\pi: \,s_t \to a_t$ such that the value function is maximized. As shown in Fig.~\ref{fig9}, the quantum DDPG includes four neural networks: the policy-network, the Q-network, the target-policy-network and the target-Q-network. Specifically, the policy NN accesses $s_t$, and the Q-NN accesses $s_t$ and $\bm\theta_t$, where $\bm\theta_t$ is the output of policy QNN. Therefore, these two QNNs cannot be executed in parallel. The target policy NN accesses $s_{t+1}$, and the target Q-NN accesses $s_{t+1}$ and $\bm\theta_{t+1}$, so these two target networks and the two original networks can be executed in parallel. In addition, the Q-network is used to approximate the value function, while the policy-network is used to approximate the policy function. 

The DDPG algorithm uses the same tricks as DQN to stabilize the training. The update of the policy-network is achieved by reducing the loss function $L(\bm{\eta})=-Q(s_t,a_t;\bm{\eta})$. Similar to DQN, the update of the Q-network in DDPG is achieved through reducing the value of the loss function $L(\bm{\omega})=E[((r_t+\bm{\gamma} \max_{a_{t+1}}Q'(s_{t+1},a_{t+1};{\bm{\omega}}'))-Q(s_t,a_t;\bm{\omega}))^2]$. Through training, the estimated value output of the Q-network will be more accurate, and the action given by the policy-network will make the Q-value higher.

\section{Quantum Reinforcement Learning in Discrete Action Space}

\begin{figure}
\centering
\includegraphics[width = 2in]{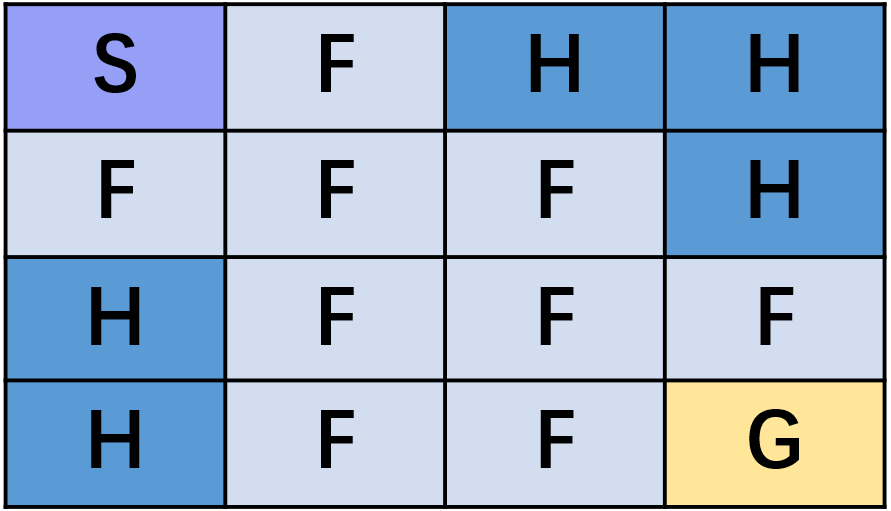}
\caption{Frozen Lake environment model. S is the starting position; F is the walkable position; H is the hole position, and G is the goal position. The actions that the agent can choose at each position are up, down, left, and right.}
\label{fig8}
\end{figure}

Besides DAS problems, our QRL framework can solve DAS problems as well. For DAS tasks, we can use quantum Q-learning or quantum DQN algorithm. Specifically, we consider a Frozen Lake environment model~\cite{brockman2016openai} in which both the action space and the state space are finite dimensional, as shown in Fig.~\ref{fig8}. In this environment, the agent can move from a grid to its neighboring grids and the goal is to move from the starting position to the target position. Some positions of the grids are walkable, while the others will make the agent fall into the water, resulting in a large negative reward, and a termination of the episode. 

In order to solve the Frozen Lake problem using our QRL framework, we number the $N$ grids from $0$ to $N-1$, and encode them into the set of basis states $S=\{\ket{j}\}$, $j=0,\cdots, j-1$ of an $N$-dimensional quantum environment, composed of $n=\log N$ qubits. At each step $t$, the state of the environment $\ket{s_t}$ equals to one of the basis states in $S$. The action $a(\bm\theta_t)$ can be represented as a parameterized action unitary $U_a (\bm\theta_t)$ on $\ket{s_{t}}$, where $\bm\theta_t$ is the action parameter. Assuming that at the position $\ket j$, there are four actions(up, down, left, and right) the agent can choose from, corresponding to the discrete set $\{U_a(\bm\theta_{j,k})\}=\{U_a(\bm\theta_{j,1}),U_a(\bm\theta_{j,2}),U_a(\bm\theta_{j,3}),U_a(\bm\theta_{j,4})\}$ where $\bm\theta_{j,k}=(\theta_{j,k}^{(1)},\cdots,\theta_{j,k}^{(n)})^T$ is a real vector. Specifically, we construct $U_a(\bm\theta_{j,k})$ as $U_a(\bm\theta_{j,k})\equiv R_y(\theta_{j,k}^{(1)})\otimes \dots \otimes R_y(\theta_{j,k}^{(n)})$, where $R_y(\theta_{j,k}^{(n)})=\exp(-i\theta_{j,k}^{(n)} \sigma_y/2)$, and $\theta^{(j)}_{k,n}\in\{0,\pi\}$. For example, in Fig.~\ref{fig8}, if the agent is at the position $\ket 5$ and wants to move right to the position $\ket 6$, then the required action parameter is $\bm\theta_{5,4}=(0,0,\pi,\pi)$, corresponding to $U(\bm\theta_{5,4})\equiv R_y(0)\otimes R_y(0)\otimes R_y(\pi)\otimes R_y(\pi)$, satisfying $U(\bm\theta_{5,4})\ket{0101}=\ket{0110}$.

We generate the reward function $f$ by introducing a reward register $\ket{r_t}$ and design the reward unitary $U_r=\sum_{j\in F} \ket j \bra j\otimes I\otimes I+\sum_{j\in H }\ket j \bra j\otimes I\otimes \sigma_x +\sum_{j\in G} \ket j \bra j\otimes \sigma_x \otimes I$ and the measurement observable $M=\sum_{j=0}^N j\ket j \bra j$. 
Then the reward $r_{t+1}$ is
\begin{align}\label{eq3}	
    r_{t+1} =f(p_{t+1}) =
    \begin{cases} 
    -1,  & p_{t+1}=0 \\
    -10,  & p_{t+1}=1 \\
    10, & p_{t+1}=2
    \end{cases}
    \end{align}
where $p_{t+1} \equiv \bra{s_t}\bra{0} U_a^\dagger(\bm{\theta}_{t,k})U^\dagger_r M U_r U_a(\bm{\theta}_{t,k}) \ket{0}\ket{s_t}$ and $r_{t+1} =f(p_{t+1})$. Here, $r_{t+1}$ is the reward for the action $U_a(\bm{\theta}_{t,k})$ at the state $\ket{s_t}$ and $\ket{0}$ is the initial state of the reward register.
With all RL elements represented as the components of a quantum circuit, we can use the quantum DQN algorithm to solve the Frozen Lake problem. In stage 1, we train the agent to find a state-action sequence to maximize the cumulative reward. In the training, the data $(\ket{s_t},a(\bm\theta),r_{t+1}) $ obtained from each interaction between the agent and the environment is recorded and these data are used to update the Q-value. In stage 2, we use the optimal policy to generate $\{U_a(\bm\theta_{0,k}),\cdots,U_a(\bm\theta_{T,k})\}$ to complete the task. In simulation, we set the size of  the replay buffer is set as 2000, the size of the  batch is set as 32, and the other parameters are  set as $\gamma = 0.9$, $\tau = 0.001$. The  quantum registers to implement the QNNs contain four qubits, and the depth of the QNN circuits is one. After 500 episodes of training, the agent can reach the target position by moving 6 steps. The sequence of actions is right, down, right, down, right, down. We can see that the agent has obtained one of the shortest paths through training.

\end{document}